\documentclass[aps,prb,amsmath,amssymb,amsfont,showpacs,superscriptaddress,reprint]{revtex4-1}
\usepackage{graphicx}
\usepackage[usenames,dvips]{color}
\definecolor{dgray}{gray}{0.3}
\definecolor{mgray}{gray}{0.5}
\definecolor{lgray}{gray}{0.7}
\usepackage{times}

\begin{document}

\title{Testing the Elliott-Yafet spin-relaxation mechanism in KC$_8$; a model system of biased graphene}

\author{G. F\'abi\'an}
\altaffiliation[Present address: ]{Department of Physics, University of Basel, Klingelbergstrasse 82, CH-4056 Basel, Switzerland}
\affiliation{Department of Physics, Budapest University of Technology and Economics and Condensed Matter Research
Group of the Hungarian Academy of Sciences, Budafoki \'{u}t 8, H-1111 Budapest, Hungary}

\author{B. D\'ora}
\affiliation{Department of Physics, Budapest University of Technology and Economics and Condensed Matter Research
Group of the Hungarian Academy of Sciences, Budafoki \'{u}t 8, H-1111 Budapest, Hungary}

\author{\'A. Antal}
\affiliation{Department of Physics, Budapest University of Technology and Economics and Condensed Matter Research
Group of the Hungarian Academy of Sciences, Budafoki \'{u}t 8, H-1111 Budapest, Hungary}
\affiliation{Condensed Matter Physics Research Group of the Hungarian Academy of Sciences, PO Box 91, H-1521 Budapest, Hungary}

\author{L. Szolnoki}
\affiliation{Department of Physics, Budapest University of Technology and Economics and Condensed Matter Research
Group of the Hungarian Academy of Sciences, Budafoki \'{u}t 8, H-1111 Budapest, Hungary}
\affiliation{Condensed Matter Physics Research Group of the Hungarian Academy of Sciences, PO Box 91, H-1521 Budapest, Hungary}



\author{L. Korecz}
\affiliation{Research Centre for Natural Sciences, Hungarian Academy of Sciences, Institute of Molecular Pharmacology, P.O. Box 17, H-1525 Budapest, Hungary}

\author{A. Rockenbauer}
\affiliation{Research Centre for Natural Sciences, Hungarian Academy of Sciences, Institute of Molecular Pharmacology, P.O. Box 17, H-1525 Budapest, Hungary}

\author{N. M. Nemes}
\affiliation{GFMC, Departamento F\'isica Aplicada III, Universidad Complutense de Madrid, Campus Moncloa, 28040 Madrid, Spain}

\author{L. Forr\'o}
\affiliation{Institute of Physics of Complex Matter, FBS Swiss Federal Institute of Technology (EPFL), CH-1015 Lausanne, Switzerland}

\author{F. Simon}
\email[Corresponding author: ]{ferenc.simon@univie.ac.at}
\affiliation{Department of Physics, Budapest University of Technology and Economics and Condensed Matter Research
Group of the Hungarian Academy of Sciences, Budafoki \'{u}t 8, H-1111 Budapest, Hungary}

\date{\today}
\begin{abstract}
Temperature dependent electron spin resonance (ESR) measurements are reported on stage 1 potassium doped graphite, a model system of biased graphene. 
The ESR linewidth is nearly isotropic and although the $g$-factor has a sizeable anisotropy, its majority is shown to arise due to macroscopic magnetization. Albeit the homogeneous ESR linewidth shows an unusual, non-linear temperature dependence, it appears to be proportional to the resistivity which is a quadratic function of the temperature. These observations suggests the validity of the Elliott-Yafet relaxation mechanism in KC$_8$ and allows to place KC$_8$ on the empirical Beuneu-Monod plot among ordinary elemental metals.
\end{abstract}

\maketitle
\section{Introduction}
Information storage and processing using electron spins, commonly referred to as spintronics \cite{FabianRMP}, is an actively studied field.
Spintronics utilizes the prolonged conservation of the spin quantum number, as the spin-relaxation time ($T_1$) usually dominates over the momentum relaxation time ($\tau$) by several orders of magnitude. In metals with inversion symmetry, $T_1$ is described by the so-called Elliott-Yafet (EY) \cite{Elliott,YafetReview} theory. The EY theory describes that the otherwise pure spin-up and down states of the conduction band are admixed due to spin-orbit coupling (SOC). The strength of the mixing is given by $\frac{L}{\Delta}$, where $L$ is the matrix element of the SOC for the conduction and a near lying band with an energy separation of $\Delta$. The conduction spin states being admixed, spin-flip transitions are possible with a low probability whenever momentum scattering occurs. Through first-order time dependent perturbation theory, Elliott connected the $g$-factor shift, the spin and the momentum relaxation times as:

\begin{gather}
\frac{1}{T_1}=\alpha_1 \left( \frac{L}{\Delta}\right)^2 \frac{1}{\tau}, \text{ and}
\label{elliottrel1}
\\
\Delta g=g-g_0=\alpha_2 \frac{L}{\Delta},
\label{elliottrel2}
\end{gather}
\noindent where $g_0=2.0023$ is the free electron $g$-factor, $\alpha_{1,2}$ are constants around unity and are determined by the band structure. The former of these equations is known as the \emph{Elliott-relation}.
The EY theory \cite{Elliott} explained spin-relaxation for most monovalent elemental metals \cite{BeuneuMonodPRB1978,MonodBeuneuPRB1979} and later studies showed its validity in one-dimensional \cite{ForroSSC1982} and polyvalent \cite{FabianPRL1998,FabianPRL1999} metals and its generalization explained the spin-relaxation in metals with strong correlations \cite{SimonPRL2008,DoraPRL2009}.

The recent discovery of graphene\cite{NovoselovSCI2004} directed the attention of spintronics research towards carbon nanostructures. The weak SOC of carbon atoms and its large mobility\cite{Bolotin2008SSC} make graphene a viable candidate for future spintronics applications, as demonstrated in non-local spin valve experiments\cite{TombrosNAT2007,Fuhrer2007APL}. Spin transport studies yield $T_1$ either by measuring the spin diffusion length $\lambda_{\textrm s} = 1/\sqrt{d} \cdot v_{\textrm F} \sqrt{T_1 \tau}$ (where $d=2,3$ is the dimension) or by a Hanle spin-precession experiment\cite{TombrosNAT2007}. The first spin transport studies indicated a very short $T_1 \approx 100\,\text{ps}$ spin-relaxation time which would be prohibitive for applications. We note that recent spin-transport experiments found a longer $T_1 \approx 2-6\,\text{ns}$ for bilayer graphene which approaches the limit of applicability \cite{GuntherodtBilayer,KawakamiBilayer}. However, the experimental situation and the value of $T_1$ is debated \cite{KawakamiPRL2010}, as well as the appropriate theoretical framework for the description of the experiments \cite{HuertasPRB2006,FabianPRB2009a,CastroNetoGrapheneSO,DoraEPL2010}.

An alternative to measure the spin-relaxation time is conduction electron spin resonance (ESR) \cite{FeherKip}. This yields $T_1$ directly from the homogeneous ESR linewidth, $\Delta B$ through $T_1=1/\gamma \Delta B$, where $\gamma/2 \pi=28.0\,\textrm{GHz/T}$ is the electron gyromagnetic ratio. An advantage of the ESR method is its contactless nature, which also allows studying powder and air sensitive materials. It is however limited by the relatively large amount of samples required, which has yet hindered ESR studies on graphene \cite{ForroPSSB2009}.

It was recently shown by angle resolved photoemission studies that alkali intercalated graphite is a model system of biased graphene since the alkali atoms effectively decouple the graphene layers \cite{GrueneisPRB2009a,GrueneisPRB2009b,VallaPRL2011}. Graphite intercalation compounds (GICs) have been known for decades \cite{DresselhausAP2002} and also, ESR was reported on them \cite{LauginiePhysB1980} but to our knowledge no thorough study has been performed in order to unravel the relaxation mechanism and in particular to study its anisotropy. Progress in the field of spin-relaxation in metals now allows a characterization using the empirical Beuneu-Monod plot \cite{BeuneuMonodPRB1978,MonodBeuneuPRB1979}, which tests the Elliott-relation (Eqs. \eqref{elliottrel1} and \eqref{elliottrel2}) based on empirical measurables alone. ESR studies on the GICs might also provide clues about the relaxation mechanism in biased graphene.

Here, we report temperature dependent conduction electron spin resonance measurements on KC$_8$ compounds made of powder and HOPG (highly oriented pyrolitic graphite) samples. KC$_8$ is the so-called stage 1 compound with the highest available K intercalation level. We find that the momentum and spin relaxation times follow a similar temperature dependence, even if they are non-linear at high temperature. This proves that the Elliott-Yafet spin mechanism is valid in KC$_8$ and allows to place this material on the Beuneu-Monod plot. The measurement on the HOPG sample shows a sizeable $g$-factor anisotropy, however it is mostly related to the anisotropy of the macroscopic susceptibility and the intrinsic $g$-factor is nearly isotropic.

\section{Experimental}

Stage 1 potassium doped graphite samples were prepared from round discs of HOPG (3 mm diameter, 50-70 $\mu$m thickness, mosaic angle: $0.4 ^{\circ}\pm 0.1^{\circ}$, ``Grade SPI-1 HOPG'', SPI Supplies Inc.) and graphite powder (grain size: 5-20 $\mu$m, Fisher Scientific Inc.). Prior to intercalation, graphite samples were annealed at $500\,^{\circ}\mathrm{C}$ in vacuum. Afterwards, samples were handled in a Ar-filled glove box to avoid exposure to oxygen and water. Doping was achieved through the two-zone vapor transport intercalation method \cite{NixonParry} at $250\,^{\circ}\mathrm{C}$ for 20 hours with a temperature gradient of $5\,^{\circ}\mathrm{C}$. The color change of HOPG from gray to gold attests a successful doping to the KC$_8$ stoichiometry. No such significant color change is apparent for the powder samples probably due to surface roughness. The resulting samples were transferred to an ESR quartz tube and sealed under 20 mbar He for the measurements \footnote{The brief exposure to the glove box sometimes induces slight surface dedoping, changing the color of the samples from gold to red, however this did not induce detectable changes in the ESR properties.}. Powder samples were mixed with dilute Mn:MgO (1.5 ppm) to allow efficient microwave penetration. The ESR-silent and doping insensitive MgO separates the graphite grains, while Mn$^{2+}$ has the added benefit of being a $g$-factor standard. 
Its $g$-factor is $g(\text{Mn}^{2+})$=2.0014 \cite{AbragamBleaneyBook} and second order hyperfine interaction effects of the Mn$^{2+}$ (Ref. \onlinecite{Torosyan}) were taken into account to determine the $g$-factor of our samples.

Experiments at 9 GHz (X band) were carried out on two commercial ESR spectrometers covering the 50-500 K temperature range with a typical microwave power of 10 mW and modulation of 0.1-0.2 mT.

\section{Results and discussion}

\subsection{The ESR lineshape}

\begin{figure}[htp]
\begin{center}
\includegraphics{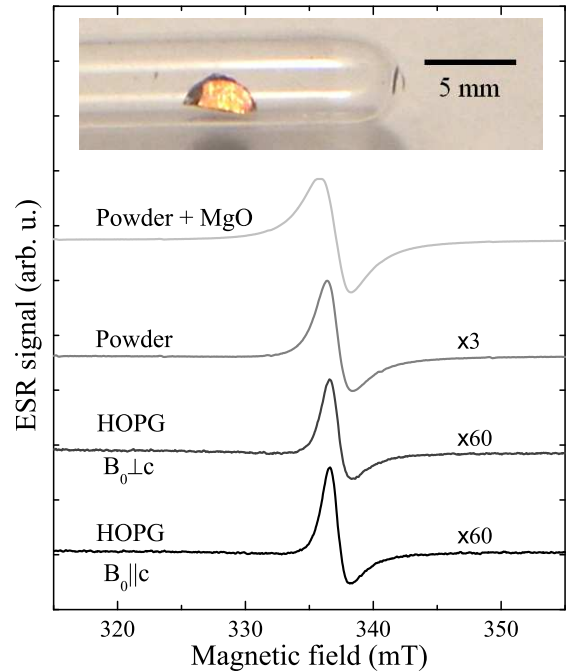}
\caption{(Color online) X-band ESR spectra of KC$_8$ for different sample forms. Note the asymmetric lineshapes characteristic for conductive samples. Inset shows a gold colored KC$_8$ sample prepared from HOPG inside a quartz tube.}
\label{spectra}
\end{center}
\end{figure}

The X-band spectra of the stage 1 potassium doped graphite in different sample forms are shown in Fig.~\ref{spectra} along with a photograph of the sample inside a quartz tube. Asymmetric derivative Loretzian lines, known as Dysonian lineshapes\cite{FeherKip,Dyson} are observed. The appearance of the Dysonian lineshapes is due to two effects. First, penetration of the exciting microwaves, which is characterized by the penetration (or \textit{skin}) depth ($\delta$) can be smaller than the sample size ($d$) for well conducting samples. Second, conduction electrons are mobile and diffuse while carrying along their spins even into regions of the sample which are not accessible for the microwave excitation. The diffusion is characterized by $T_{\text{D}}$, which is the time it takes for an electron to diffuse through the penetration depth. The Dysonian lineshape takes a number of different forms depending on the values of $\delta$, $d$, $T_1$, and $T_{\text{D}}$.

We found that the ESR data can be fitted with a mixture of Lorentzian absorption and dispersion curves with an almost 1:1 ratio \cite{WalmsleyJMR1996}. This situation is the so-called ``NMR limit'' of the Dysonian lineshape and is described by Eq. (3.7) in the work of Feh\'{e}r and Kip \cite{FeherKip}. This occurs when $T_{\text{D}} \>>T_1$ i.e. the electrons are diffusing through the penetration depth very slowly. This behavior in doped graphite was explained by Walmsley \textit{et al.} for n-doped GICs [Ref. \onlinecite{WalmsleySynMet1989}]. The argument considered the conduction anisotropy of graphite, i.e. $\rho_{c}\gg \rho_{ab}$ and the platelet-like structure of HOPG such that sample dimensions in the $ab$ crystalline plane are much larger than that along the $c$-axis. The conduction anisotropy and the good $ab$ plane conductivity results in a relatively small penetration depth. However, somewhat counterintuitively, the relevant diffusion term, which characterizes $T_{\text{D}}$ is that along the $c$-axis. The diffusion being limited along this direction, the $T_{\text{D}}\gg T_1$ holds which explains the experimental observation.

The fact that we observe a similar lineshape for both the doped HOPG and graphite powder samples indicates that this situation also holds for the latter. For all spectra, the asymmetry is moderate enough to fit the acquired spectra with such Lorentzian combination curves. The parameters of these fits yield the intensity, resonance field, and linewidth (HWHM: half width at half maximum of the Lorentzian resonance curves).



\subsection{The temperature dependent ESR linewidth}

\begin{figure}[htp]
\begin{center}
\includegraphics[width=3in]{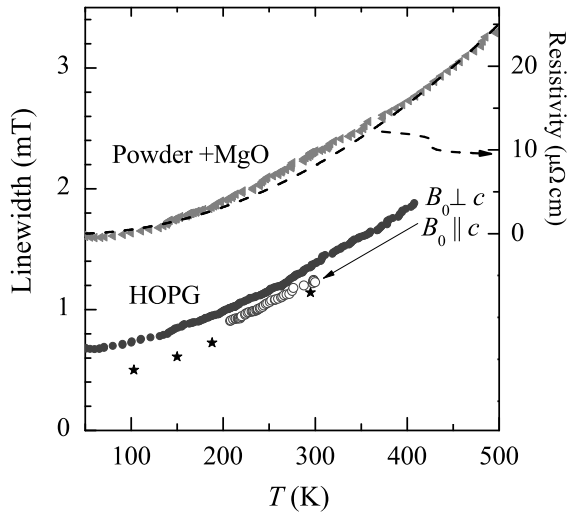}
\caption{Temperature dependence of the ESR linewidth in KC$_8$ (left axis) at 9 GHz. HOPG: $B_0$ $\perp$ $c$ (\textcolor{dgray}{$\bullet$}) and $B_0$ $\parallel$ $c$ ({$\circ$}), powder (\textcolor{mgray}{$\blacktriangleleft$}). Data on HOPG with $B_0$ $\parallel$ $c$ ($\bigstar$) from Ref.~\onlinecite{LauginiePhysB1980} is shown for comparison. Note the small linewidth anisotropy as found for the HOPG sample. Dashed curve shows in-plane resistivity data (right axis) from Ref.~\onlinecite{PotterSSC1981}, with an axis offset in order to shows that it follows the same temperature dependence as the linewidth.}
\label{w}
\end{center}
\end{figure}

The temperature dependence of the ESR spectra was measured for the different sample forms. The extracted linewidths are shown in Fig. \ref{w}. Doped HOPG samples were studied only up to 450 K as for higher temperatures the linewidth changes irreversibly, possibly due to the evaporation of potassium atoms from the surface. The anisotropy of the linewidth was also studied, for an external magnetic field, $B_0$, parallel ($B_0 \| c$) or perpendicular ($B_0 \perp c$) to the $c$-axis.

The temperature dependence of the linewidth shows the same trend irrespective of the sample form. The curves mainly differ by a constant linewidth term: the powder sample has a characteristically larger linewidth, which is either due to the larger effective surface which gives rise to additional relaxation \cite{Dyson} or due to a higher concentration of impurities. In general, the linewidth can be written in analogy to the Matthiessen's rule for resistivity, i.e. as a sum of different contributions:

\begin{gather}
\Delta B(T)=\Delta B_{\textrm {hom}}(T)+\Delta B_{\textrm {hom},0}
\label{LinewidthGeneral}
\end{gather}

\noindent Here, $\Delta B_{\textrm{hom},0}$ is a temperature independent (i.e. residual) homogeneous relaxation such as caused by the surface (or impurities) and $\Delta B_{\textrm{hom}}(T)$ is the temperature dependent homogeneous relaxation, which is in the focus herein.

The linewidth data for the HOPG in the two orientations show a tiny anisotropy. The linewidth data for the HOPG are in good agreement with that reported by Lauginie \textit{et al.} in Ref.~\onlinecite{LauginiePhysB1980}, which means that not only the temperature dependent $\Delta B_{\textrm {hom}}(T)$ is the same but there is a generic $\Delta B_{\textrm {hom},0}$ linewidth which is the same for HOPG samples from different sources.

Our temperature resolution allows to observe a non-linearity in the temperature dependence of the linewidth which has been not reported, yet. It is best shown by a direct comparison to the temperature dependent in-plane resistivity, $\rho(T)$ as it is done in Fig. \ref{w}. The two kinds of data are scaled together and are shown with an offset since the residual and temperature dependent contributions to $\Delta B(T)$ and $\rho(T)$ do not necessarily scale together.

The temperature dependence of the in-plane resistivity, $\rho(T)$, is quadratic \cite{SuematsuJPSJ1980,PotterSSC1981} and is shown in Fig.~\ref{w}. from Ref.~\onlinecite{PotterSSC1981}. For usual metals, $\rho(T)$ is linear for temperatures $T \gtrsim \Theta_{\text{D}}/2$. The Debye temperature is $\Theta_{\text{D}}=235\,\text{K}$ in KC$_{8}$ \cite{DresselhausAP2002}, whose value underlines the unusual nonlinearity of $\rho(T)$ in KC$_8$. It was argued that this surprising quadratic temperature dependence of $\rho(T)$ in KC$_{8}$ arises from electron-electron interaction \cite{SuematsuJPSJ1980}. We demonstrate in Fig. \ref{w}., that the temperature dependence of the ESR linewidth follows closely that of the resistivity. This leads us to suggest that the Elliott-Yafet theory is the appropriate description of the spin-relaxation in KC$_8$.

\subsection{The $g$-factor in KC$_8$}

\begin{table}[t]
\caption{$g$-factor and in its shift with respect to $g_0=2.0023$ for the different KC$_8$ compounds. The procedure to obtain the corrected $g$-factor is explained in the text. Results from Ref. \onlinecite{LauginiePhysB1980} are also shown. Errors are estimated from the variance of the data for different samples.}
\begin{tabular*}{0.9\columnwidth}{@{\extracolsep{\fill}}lccc}
\hline \hline
Host compound& Orientation& $g$ & $ \Delta g \times 10^{4}$\\
\hline
HOPG (measured)&$B_0 \parallel c$&$2.0013$ & $-10\pm 	1$\\
HOPG (measured)&$B_0 \perp c$&$2.0028$ & $5\pm1$\\
HOPG (corrected)&$B_0 \parallel c$&$2.0030$ & $7\pm 	1$\\
HOPG (corrected)&$B_0 \perp c$&$2.0028$ & $5\pm 1$\\
\hline
powder(measured)& & $2.0024$ &$1\pm 1$\\
powder(calculated)& & $2.0024$&\\
\hline
HOPG (Ref.~\onlinecite{LauginiePhysB1980}.) &$B_0 \parallel c$&$2.0016$ & $- 7$ \\
HOPG (Ref.~\onlinecite{LauginiePhysB1980}.) &$B_0 \perp c$&$2.0030$ & $7$ \\
\hline
\end{tabular*}
\label{gfactor}
\end{table}

In the following, we discuss the value of the $g$-factor in KC$_8$, which allows to follow the procedure of Beuneu and Monod \cite{BeuneuMonodPRB1978} in verifying the validity of the Elliott-relation in KC$_8$ \emph{quantitatively} (Eqs. \eqref{elliottrel1} and \eqref{elliottrel2}). Room temperature $g$-factors were obtained by comparison to the Mn$^{2+}$ reference and the results are summarized in Table \ref{gfactor} along with data from previous studies \cite{LauginiePhysB1980}. While the present data for the HOPG show the same anisotropy as in the literature, an overall difference of $2-3 \cdot 10^{-4}$ is observed for both orientations, whose origin is unclear \footnote{It may be related to the different determination of the $g$-factor; in the previous work the $g$-factor was determined graphically \cite{LauginiePhysB1980}}. We checked the consistency of the $g$-factor values for the HOPG and the powder samples with a numerical comparison: the anisotropic HOPG $g$-factor data was used to simulate powder spectra with a uniaxial $g$-factor anisotropy, which was then fitted with a Lorentzian. It yielded $\Delta g=1\cdot 10^{-4}$, which agrees with the experimental data for the powder as shown in Table \ref{gfactor}.

A correction is required to determine the intrinsic $g$-factor in KC$_8$ due to its macroscopic magnetism. Again, usual alkali metals (which are the textbook example of the EY theory) do not have a sizeable macroscopic magnetism. Graphite has unusual magnetic properties, as it has a relatively large and anisotropic diamagnetic susceptibility. This magnetism was associated with the orbital currents in graphite \cite{HuberPRB2004} and it not only affects the macroscopic properties but it also couples to the microscopic measurables. GICs also exhibit a sizeable macroscopic susceptibility but it is paramagnetic and its absolute value is smaller than that of pristine graphite. Such a magnetism is known in general to interact locally with the conduction electrons \cite{BarnesAdvPhys}, thus the measured $g$-factor, $g_\text{meas}$, is different from the intrinsic one, $g_\text{intr}$. This situation is well known in NMR, where local magnetism, e.g. those due to orbital currents couple to the nuclei differently than the macroscopic magnetism \cite{AbragamBook}.

The resulting $g$-factor in KC$_8$ is determined by the static susceptibility, $\chi_0$, and a mean-field like coupling constant $A$ and reads:

\begin{equation}
g_\text{meas}=g_\text{intr}\left(1+A \chi_0\right),
\label{gcorr}
\end{equation}

\begin{figure}[htp]
\begin{center}
\includegraphics{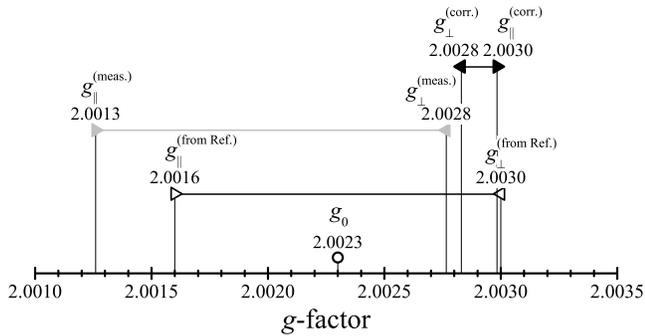}
\caption{Illustration of $g$-factors in KC$_8$: the $g_0=2.0023$ free electron $g$-factor ($\bigcirc$), data from Ref.~\onlinecite{LauginiePhysB1980} ({$\vartriangleleft$} and {$\vartriangleright$}), current measurements (\textcolor{lgray}{$\blacktriangleleft$} and \textcolor{lgray}{$\blacktriangleright$}), and current measurements corrected by coupling to macroscopic susceptibility ($\blacktriangleleft$ and $\blacktriangleright$).}
\label{Fig3_gfactor}
\end{center}
\end{figure}

We require some simplifying assumptions to determine the intrinsic $g$-factors in KC$_8$: the coupling constant $A$ is isotropic and is not affected by the alkali doping. We obtain $A=-1.12\cdot 10^3$ $\frac{\text{g}}{\text{emu}}$ from $\chi_0^\text{graphite}(B_0 \parallel c) = -21.1\cdot 10^{-6}\,\frac{\text{emu}}{\text{g}}$ (Ref.~\onlinecite{DiSalvoPRB1979}) and $\Delta g^\text{graphite} (B_0 \parallel c) = +0.0473$ (Ref. \onlinecite{WagonerPR1960}).
Here, we assumed that the whole $g$-factor shift in pristine graphite stems from the diamagnetic susceptibility and that its intrinsic $g$-factor is $g_0$.

KC$_8$ has macroscopic susceptibilities of $\chi_0^\text{KC8}(B_0 \parallel c)=1.02\cdot 10^{-6}\,\frac{\text{emu}}{\text{g}}$ and $\chi_0^\text{KC8}(B_0 \perp c)=0.28\cdot 10^{-6}\,\frac{\text{emu}}{\text{g}}$ (Ref.~\onlinecite{DiSalvoPRB1979}). When these values are corrected by the diamagnetic core shielding and the Pauli term stemming from conduction electrons \cite{DresselhausAP2002}, the true contributions from orbital magnetism is obtained: $\chi_\text{orb}^\text{KC8}(B_0 \parallel c)=0.769 \cdot 10^{-6}\,\frac{\text{emu}}{\text{g}}$ and $\chi_\text{orb}^\text{KC8}(B_0 \perp c)=0.029\cdot 10^{-6}\,\frac{\text{emu}}{\text{g}}$. These values lead to the corrected, intrinsic $g$-factors in Table \ref{gfactor}, which are both positive in sign and have a much lower anisotropy than the non-corrected values. The small $g$-factor anisotropy is in agreement with the similarly small anisotropy of the ESR linewidth.

\subsection{The microscopic theory of spin-relaxation in KC$_8$}

The magnitude of anisotropy is important for the microscopic theory of spin-relaxation in intercalated graphite.
This is also thought to be relevant for graphene as the GICs were suggested to act as its model system \cite{GrueneisPRB2009a,GrueneisPRB2009b,VallaPRL2011}; namely it was shown that the conduction band in KC$_8$ consists of Dirac cones (such as in graphene) with a Fermi surface separated by 1.35 eV from the Dirac point \cite{GrueneisPRB2009b}.
Therefore, we attempt to adopt the description which was used in biased graphene to explain the spin-relaxation behavior in KC$_8$.

The kinetic energy in graphene is given by the low-energy Dirac Hamiltonian, which is obtained by expanding the tight-binding Hamiltonian around the corners of the hexagonal Brillouin zone (the $K$ and $K'$ points), as \cite{GeimRMP}
\begin{equation}
H_\text{Dirac}=v_F\left(\sigma_x p_x+\sigma_y \tau_z p_y\right)-\mu,
\label{Dirac}
\end{equation}

\noindent where $\sigma$'s describe the two sublattices of the graphene honeycomb lattice,
$v_F=10^6$~m/s is the Fermi velocity, $\tau_z=1$ (-1) describes the $K$ ($K'$) degeneracy, and the $\mu$ chemical potential accounts for a finite doping. For pristine, undoped graphene, $\mu=0$.

The resulting energy spectrum reads as a function of the momentum ${\bf p}$ as $\epsilon({\bf p})=v_F|{\bf p}|$.
This has to be supplemented with a spin-orbit coupling term to explain the spin-relaxation. It was argued previously that the intrinsic, i.e.~atomic spin-orbit coupling is responsible for the spin-relaxation and $g$-factor shift
for both graphene and the GICs \cite{DoraEPL2010} provided the linear energy dispersion applies. The spin-orbit Hamiltonian is:
\begin{equation}
H_{\textrm{SO}}=\Delta_{\textrm{intr}} {\tau}_z {\sigma}_z {s}_z,
\label{SOHamiltionian}
\end{equation}

\noindent where $\Delta_{\textrm{intr}}$ is the strength of the intrinsic SOC, ${s}_z$ accounts for the physical spin, and ${\sigma}_z$ measures the energy imbalance between the two sublattices. The above form of $H_{\textrm{SO}}$ satisfies time-reversal invariance, as expected.

Eqs. \eqref{Dirac} and \eqref{SOHamiltionian} predict a significant anisotropy: the
spin-relaxation would be sizeable when $B \perp c$ and it would vanish for $B \parallel c$, since only $s_z$ is involved in the above Hamiltonians.
On the other hand, the $g$-factor shift would be finite when $B \parallel c$ and zero when $B \perp c$. As mentioned, the experimental data does not show a sizeable anisotropy for either of these parameters. At present, we do not have a consistent explanation for the difference between the theoretical model and the experimental data. A possibility is that the simplest linear band model is not sufficient and that additional bands and spin-orbit couplings are involved such as e.g. those of the K$^{+}$. Although these bands are well separated from the conduction band \cite{DresselhausAP2002}, they might influence the spin-relaxation properties due to a weak hybridization of the alkali and graphite bands. A similar effect was invoked to explain the alkali dependent linewidth in alkali doped fullerides \cite{DoraPRL2009}.

We note, that the lack of such extreme anisotropy was also observed in graphene spin transport measurements \cite{TombrosPRL2008}. Interestingly,
even the magnitude and the sign of the present $\sim$ 25 \% linewidth anisotropy agrees with the $\sim 20$ \% value observed by
Tombros \textit{et al.} in Ref.~\onlinecite{TombrosPRL2008}.

\subsection{The spin-relaxation mechanism in KC$_8$ and the Beuneu-Monod plot}

\begin{figure}[htp]
\begin{center}
\includegraphics{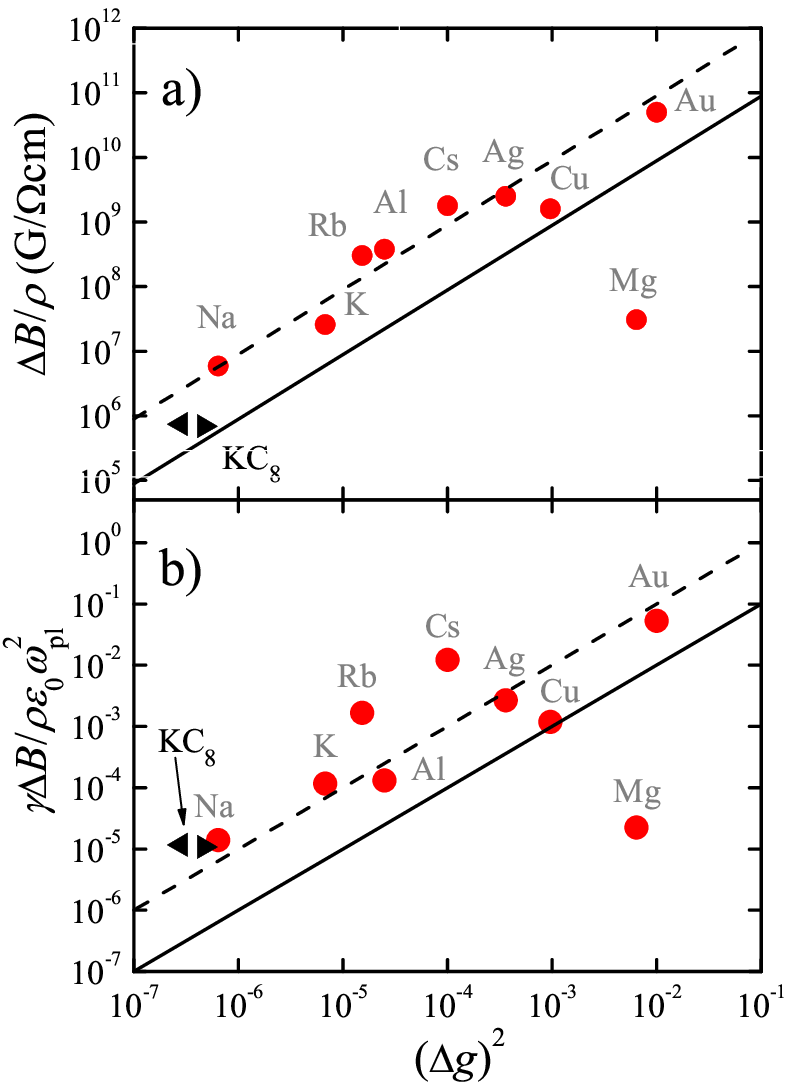}
\caption{The Beuneu-Monod plot showing the connection between $(\Delta g)^2$ and ratio of the linewidth and resistivity ($\Delta B / \rho$) for pure metals (\textcolor{red}{$\bullet$}) from Ref.~\onlinecite{BeuneuMonodPRB1978} and for KC$_8$ for $\perp$ ($\blacktriangleleft$) and $\parallel$ ($\blacktriangleright$) orientations (error bars are smaller than the symbols). a) The original version with cgs units, b) the corrected plot with the dimensionless reduced linewidth ($\gamma \Delta B / \rho \varepsilon_0 \omega_\text{pl}^2$) and considering the variation of $\omega_{\textrm{pl}}$ among the materials. Solid lines correspond to $\alpha_1/\alpha^2_2=1$ and the dashed curve is the best fitting $\alpha_1/\alpha^2_2=10$ as found in Ref.~\onlinecite{BeuneuMonodPRB1978}.}
\label{Fig4_BM_test_sum}
\end{center}
\end{figure}

The above discussed temperature dependent $\Delta B_{\textrm {hom}}(T)$, the $\rho(T)$ data from the literature (Ref.~\onlinecite{PotterSSC1981}) (both are free from the temperature independent terms) and the corrected value for $\Delta g$ allows us to discuss the validity of the Elliott-relation in KC$_8$. The Elliott-relation can be rewritten in terms of these experimental measurables as:
\begin{equation}
\frac{\Delta B}{\rho}=\frac{\varepsilon_0 \omega_\text{pl}^2}{\gamma}\frac{\tau}{T_1}=\frac{\varepsilon_0 \omega_\text{pl}^2}{\gamma}\frac{\alpha_1}{\alpha_2^2}(\Delta g)^2
\label{BMratio}
\end{equation}

\noindent where $\varepsilon_0$ is the dielectric constant, $\omega_\text{pl}$ is the plasma frequency. In the summarizing work of Beuneu and Monod \cite{BeuneuMonodPRB1978} for most metals, $\frac{\Delta B}{\rho}$ ratio was found to be linearly proportional to $(\Delta g)^2$ with a constant coefficient of $10^{11}\,\text{G}/\Omega\text{cm}$ (when using cgs units), neglecting the variation of $\omega_\text{pl}^2$ from metal to metal.

The original Beuneu-Monod plot is shown in Fig.~\ref{Fig4_BM_test_sum}a. along with the present data for KC$_8$, calculated from the present $\Delta B_{\textrm {hom}}(T)$ and $\Delta g$ data and from the $\rho(T)$ in Ref.~\onlinecite{PotterSSC1981}. Clearly, the data points for both HOPG orientations lie at an order of magnitude lower value than the majority of the other materials. This can be explained by the relatively low value of $\omega_\text{pl}=2.35\,\text{eV}$ in KC$_8$ \cite{DresselhausAP2002} as compared to the usual value of $> 5\,\text{eV}$ in elemental metals. In the original work of Beuneu and Monod, variation of $\omega_\text{pl}$ among the metals was not taken into account. We note that Petit \textit{et al.} also suggested previously \cite{PetitPRB1996} that the lower $\omega_\text{pl}$ explains why data for alkali doped fullerides also fall relatively low on the Beuneu-Monod plot.

In Fig.~\ref{Fig4_BM_test_sum}b., we show the corrected data by introducing the dimensionless reduced linewidth, $\gamma \Delta B / \rho \varepsilon_0 \omega_\text{pl}^2$, as a function of $(\Delta g)^2$. With this correction, the data for KC$_8$ lie on the $\alpha_1/\alpha^2_2=10$ straight line, which was found to best fit the experimental data for most elemental metals in Ref.~\onlinecite{BeuneuMonodPRB1978}. We note that with this correction, the data for Rb and Cs does not seem to agree with this straight line in contrast to the original assertion. Whether Rb and Cs represents an anomalous situation herein or the somewhat old experimental results need to be revisited, requires additional work. Nevertheless, the present result confirms that KC$_8$ also follows the Elliott-Yafet theory of spin-relaxation and even the $\alpha_{1,2}$ parameters, which are sensitive to the band structure, are similar to that in ordinary elemental metals.

\section{Conclusions}

In summary, we found the Elliott-Yafet theory of spin-relaxation to be valid for the KC$_8$ stage 1 graphite intercalation compound, with the proportionality of the homogeneous linewidth and in-plane resistivity in agreement with the $g$-factor shifts as for most elemental metals. It remains, however open for further investigations whether the result can be applied directly for the spin-relaxation in biased graphene as expected based on the band structure calculations and ARPES studies \cite{GrueneisPRB2009a,GrueneisPRB2009b,VallaPRL2011}.

\begin{acknowledgments}
The Authors dedicate this work to the memory of John E. Fischer. We thank A. J\'{a}nossy and T. Feh\'{e}r for enlightening discussions. Work supported by the ERC Starting Grant Nr. ERC-259374-Sylo, by the Hungarian State Grants (OTKA) Nr.~K72613, CNK80991, K73361, and K68807, and by the New Sz\'{e}chenyi Plan Nr. T\'{A}MOP-4.2.2.B-10/1--2010-0009. BD acknowledges the Bolyai programme of the Hungarian Academy of Sciences. The Swiss NSF is acknowledged for support. NMN acknowleges the ``Ramon y Cajal'' contract of the Spanish MICINN.
\end{acknowledgments}


%

\end{document}